\documentclass[jcp,graphicx, twocolumn]{revtex4-1}

\usepackage{amsmath}
\usepackage{amsfonts}
\usepackage{mathtools}
\usepackage{braket}
\usepackage{commath}

\usepackage[version=3]{mhchem}
\usepackage[left=1.5cm, right=1.5cm, top=1.785cm, bottom=2.0cm]{geometry}
\usepackage{balance}
\usepackage{mathptmx}
\usepackage{graphicx} 
\usepackage{lastpage}
\usepackage{float}
\usepackage{fancyhdr}
\usepackage{fnpos}
\usepackage[english]{babel}
\addto{\captionsenglish}{%
  \renewcommand{\refname}{Notes and references}
}
\usepackage{array}
\usepackage{droidsans}
\usepackage{charter}
\usepackage[usenames,dvipsnames]{xcolor}
\usepackage{setspace}
\usepackage[compact]{titlesec}
\usepackage{hyperref}
\usepackage{bm}
\DeclareMathAlphabet{\mathcal}{OMS}{cmsy}{m}{n}
\SetMathAlphabet{\mathcal}{bold}{OMS}{cmsy}{b}{n}

\draft 

\begin{document}


\title{Formally exact, arbitrarily scalable simulations of exciton dynamics in molecular materials} 



\author{Leonel Varvelo, Jacob K. Lynd, and Doran I. G. Bennett\textit{$^{\ast}$}}
\affiliation{Department of Chemistry, Southern Methodist University, PO Box 750314, Dallas, TX, USA. E-mail: doranb@smu.edu}


\date{\today}

\begin{abstract}
Excited state carriers, such as excitons, can diffuse on the 100 nm to micron length scale in molecular materials, but they only delocalize over short length scales due to coupling between electronic and vibrational degrees-of-freedom. Here, we leverage the locality of excitons to adaptively solve the hierarchy of pure states equations (HOPS). We demonstrate that our adaptive HOPS (adHOPS) methodology provides a formally exact and size-invariant (i.e. $\mathcal{O}(1)$) scaling algorithm for simulating mesoscale quantum dynamics. We provide proof-of-principle calculations for exciton diffusion on linear chains containing up to 1000 molecules.
\end{abstract}


\maketitle 

\section{Introduction}
New molecular materials, particularly organic semiconductors, offer remarkable and tunable functionality for photonic, opto-electronic, and light harvesting applications. The photophysical properties of molecular materials arise from diffusion of excited-state carriers (e.g., electronic excitations, called `excitons') across the 10 nm to 1 $\mu$m length scale. These mesoscale exciton dynamics are sensitive to both the molecular properties of the material building blocks and the structural heterogeneities arising on these length scales, which include everything from point defects to grain boundaries. Traditional bulk spectroscopies provide only indirect evidence for the essential role of structural heterogeneity in exciton transport. The recent development of spatially-resolved non-linear spectroscopy provides a remarkable new lens by which to study exciton dynamics in heterogeneous materials. \cite{delorImagingMaterialFunctionality2020, ginsbergSpatiallyResolvedExciton2020} Interpreting spatially-resolved spectroscopic signals, however, remains challenging due to the absence of corresponding simulations.

Simulating exciton transport dynamics in heterogeneous materials on the 10 nm - 1 $\mu$m length scale remains an outstanding theoretical challenge. Organic semiconductors often combine close intermolecular packing with correspondingly large coupling between electronic states (V) on adjacent molecules and large intramolecular electron-vibrational coupling ($\lambda$).\cite{nematiaramModelingChargeTransport2020} Perturbative equations-of-motion, such as F\"{o}rster theory, can be convenient for simulating large aggregates, but are not applicable when V and $\lambda$ are comparable in magnitude. Similarly, in the absence of a clear separation of timescales between vibrational and electronic degrees of freedom, Markovian equations-of-motion, such as Redfield theory,\cite{yangInfluencePhononsExciton2002} struggle to capture the rich dynamics of excitation transport. There are a variety of non-perturbative, non-Markovian equations-of-motion, such as multi-layer multi-configuration time-dependent Hartree (ML-MCTDH), \cite{schulzeMultilayerMulticonfigurationTimedependent2016} time-evolving density operator with orthogonal polynomials (TEDOPA),\cite{tamascelliEfficientSimulationFiniteTemperature2019} hierarchically-coupled equations of motion (HEOM),\cite{tanimuraNumericallyExactApproach2020} and quasi-adiabtic path integrals (QUAPI).\cite{makriImprovedFeynmanPropagators1991} All of these techniques, however, share an exponential scaling of computational complexity with the number of molecules. While efficient and parallelized implementations of formally exact methods have been developed -- for example, distributed memory HEOM \cite{kramerEnergyFlowPhotosystem2018, kramerEfficientCalculationOpen2018} -- the exponential scaling severely limits even high-performance simulations of molecular aggregates.

Recently, there have been a few notable developments towards highly-scalable equations-of-motion for exciton dynamics. Modular path integrals \cite{makriModularPathIntegral2018,makriCommunicationModularPath2018} provide a dramatic reduction in computational cost of QUAPI, but retain an overall linear scaling with the number of molecules and are most efficient when molecules exhibit only nearest-neighbor coupling. Dissipation-assisted matrix product factorization (DAMPF) \cite{somozaDissipationAssistedMatrixProduct2019} extends TEDOPA to efficiently describe large numbers of vibrational degrees of freedom ($>10$) on each molecule, but it maintains between a quadratic and cubic scaling with the number of molecules. For both modular path integrals and DAMPF, the residual scaling makes it challenging to apply these methods to mesoscale calculations containing thousands to millions of molecules. Indeed, any density matrix approach will suffer from residual scaling with system size at long times due to the spread of ensemble population density across molecules. 

Stochastic simulations, which decompose the ensemble into a collection of excited trajectories, 
can enable calculations on arbitrarily large molecular aggregates, even at long time. Delocalized kinetic Monte Carlo \cite{balzerDelocalisedKineticMonte2021} and the kinetic Monte Carlo version of generalized F\"{o}rster theory \cite{amarnathMultiscaleModelLight2016} are stochastic approaches that calculate the rate of transport between clusters of strongly interacting molecules and can be readily extended to mesoscale calculations. Both of these methods, however, use a perturbative approximation to partition state space and calculate rates between adjacent spatial regions. The development of a non-perturbative, non-Markovian approach for mesoscale simulations would provide an important benchmark for new equations-of-motion and could offer insight into processes with debated mechanisms, such as charge separation in organic photovoltaic materials.\cite{balzerDelocalisedKineticMonte2021, hoodEntropyDisorderEnable2016, jailaubekovHotChargetransferExcitons2013, monahanDirectObservationEntropyDriven2015} 

Here, we present a non-perturbative, non-Markovian, and arbitrarily scalable stochastic method for simulating exciton transport. First, we introduce some preliminary discussion of the Hamiltonian considered and our base equation-of-motion, the hierarchy of pure states (HOPS) \cite{suessHierarchyStochasticPure2014}. Next, we discuss locality in HOPS calculations and present an algorithm for constructing an adaptive basis. Finally, we present proof-of-concept calculations using the adaptive HOPS (adHOPS) equation-of-motion that demonstrate both its accuracy and size-invariant (i.e. $\mathcal{O}(1)$) scaling for large molecular aggregates. 

\section{Preliminaries}
\subsection{Hamiltonian}
We divide the exciton Hamiltonian into three parts
\begin{equation}
    \label{eq:Ham_Full}
    \hat{H}_{T} = \hat{H}_{\textrm{S}}\otimes \hat{\mathbb{I}}_B + \hat{H}_{\textrm{int}} + \hat{\mathbb{I}}_S\otimes \hat{H}_{\textrm{B}}
\end{equation}
where $\hat{H}_{\textrm{S}} = \sum_n \vert n \rangle E_n \langle n \vert + \sum_{n \neq m} \vert n \rangle V_{n,m} \langle m \vert$ describes the electronic system and $\hat{H}_{\textrm{B}} = \sum_{n,q} \hbar \omega_{q_n} (\hat{a}^\dagger_{q_n} \hat{a}_{q_n} + 1/2)$ represents the thermal environment arising from molecular vibrations.  The influence of coupling between the electronic system and vibrational `bath'  ($\hat{H}_{\textrm{int}} =  \sum_{n,q} \kappa_{q_n} \hat{L}_n \hat{q}_n$) can be described in terms of the system-bath coupling operators ($\hat{L}_n$) and the two-point correlation functions 
\begin{equation}
\label{eq:C_t}
\alpha_n(t) = \frac{1}{\pi}\int_0^\infty d\omega J_n(\omega) \big(\coth(\hbar \omega /2 k_B T) \cos(\omega t) - i \sin( \omega t)\big)
\end{equation}
where $T$ is the temperature and $J_n(\omega)$ is the spectral density. In the following, we assume that each pigment has an independent thermal environment that drives fluctuations in excitation energy. Said in other words, we assume that the system-bath coupling operator is a site-projection operator ($\hat{L}_n = \vert n \rangle\langle n \vert$ ). We describe the thermal environment of each pigment by a Drude-Lorentz spectral density 
\begin{equation}
J_n(\omega) = 2 \lambda_n \gamma_n \frac{\omega}{\omega^2 + \gamma_n^2}
\end{equation} 
which, at high temperature ($\gamma/k_B T < 1$), allows for a convenient exponential decomposition of the correlation function
\begin{equation}
\label{eq:alpha_dl_coarse}
    \alpha_n(t) = g_n e^{-\gamma_n t/\hbar}
\end{equation}
where $g_n = 2 \lambda_n k_B T - i \lambda_n \gamma_n$. In the following we will use $\lambda = \gamma = 50 \textrm{ cm}^{-1}$, $V=25-250\textrm{ cm}^{-1}$ and T=$295$K, which are comparable to the parameters used for many simulations of photosynthetic pigment protein complexes and are known to fall into the broad intermediate regime where perturbative approximations break down. \cite{ishizakiUnifiedTreatmentQuantum2009}

\subsection{Hierarchy of Pure States (HOPS)}
The non-Markovian quantum state diffusion (NMQSD) equation \cite{diosiNonMarkovianStochasticSchrodinger1997} decomposes the time-evolution of the reduced density matrix for the system into an ensemble average over stochastic pure states indexed by a complex stochastic processes $z_{n,t}$
\begin{equation}
\label{eq:reduced_den_linear}
    \rho_S = \mathbb{E}[\vert \psi(t; z_{n,t}) \rangle \langle \psi(t; z_{n,t}) \vert]
\end{equation}
where $\mathbb{E}[z_{n,t}] =0$,  $\mathbb{E}[z_{n,t} z_{n,s}] =0$, and $\mathbb{E}[z^*_{n,t} z_{n,s}] = \alpha_n(t-s)$. The equation-of-motion for the independent stochastic trajectories is
\begin{flalign}
\begin{aligned}
    \partial_t \vert \psi(t; z_{n,t}) \rangle = &\big (- i \hat{H}_S  + \sum_n \hat{L}_n z^*_{n,t}\big)\vert \psi(t; z_{n,t}) \rangle \\&- \sum_n  \hat{L}^\dagger_n \int_0^t ds \alpha_n(t-s) \dfrac{\delta \vert \psi(t; z_{n,t}) \rangle}{\delta z^*_{n,s}}.
\end{aligned}
\end{flalign}
The NMQSD equation is formally exact and is equivalent to solving Feynman Path Integrals with the Feynman-Vernon influence functional,\cite{diosiNonMarkovianStochasticSchrodinger1997} but the functional derivative in the last term makes direct solution of the stochastic trajectories impractical except in special cases.

The hierarchy of pure states (HOPS) equations provide a numerically tractable version of NMQSD by rewriting the functional derivative as a set of coupled differential equations.\cite{suessHierarchyStochasticPure2014} Briefly, the sum of integrals over a functional derivative in the final term of the NMQSD equation is defined as a sum of first order auxiliary wave functions:
\begin{equation}
    \vert \psi^{(\vec{e}_n)}(t; z_{n,t}) \rangle = \int_0^t ds \alpha(t-s) \dfrac{\delta \vert \psi(t; z_{n,t}) \rangle}{\delta z^*_{n,s}}
\end{equation}
giving 
\begin{flalign}
\begin{aligned}
    \partial_t \vert \psi^{(\vec{0})}(t; z_{n,t}) \rangle = &\big (- i \hat{H}_S  + \sum_n \hat{L}_n z^*_{n,t}\big)\vert \psi^{(\vec{0})}(t; z_{n,t}) \rangle \\&- \sum_n  \hat{L}^\dagger_n \vert \psi^{(\vec{e}_n)}(t; z_{n,t}) \rangle
\end{aligned}
\end{flalign}
where we have now introduced a vector label into the equations to index the different components. The physical wave function is given by $\vert \psi^{(\vec{0})}(t; z_{n,t}) \rangle$. The first order auxiliaries are indexed by unit vectors with non-zero index at their $n^{th}$ element ($\vec{e}_n$). When the correlation function $\alpha_n(t)$ is written as an exponential (or sum of exponentials), the time-evolution of the first order auxiliary wave functions $\vert \psi^{(\vec{e}_n)}(t; z_{n,t}) \rangle$ introduces the second-order auxiliary wave functions ($ \vec{e}_n + \vec{e}_m$), and so on, \textit{ad infinitum}. The resulting general expression, called the `linear HOPS equation,' is 
\begin{flalign}
\begin{aligned}
\label{eq:LinearHops}
\partial_t\psi^{(\Vec{k})}_{t}
= \big(-i\hat{H}_S - \Vec{k} \cdot \Vec{\gamma} + \sum_{n} \hat{L}_{n} z^*_{t,n}\big)&\vert \psi^{(\vec{k})}(t; z_{n,t}) \rangle 
\\+ \sum_{n} \Vec{k}[n] g_{n} \hat{L}_{n}&\vert \psi^{(\Vec{k} -\Vec{e}_{n})}(t; z_{n,t}) \rangle
\\ - \sum_{n} \hat{L}^{\dagger}_{n} & \vert \psi^{(\Vec{k}+\Vec{e}_{n})}(t; z_{n,t}) \rangle 
\end{aligned}
\end{flalign}
where we have introduced a general vector $\Vec{k}$ to index auxiliary wave functions, $\Vec{k}[n]$ is the $n^{th}$ element of the index vector, $\Vec{\gamma}$ is the vector of correlation function exponents ($\gamma_n$), and terms involving any auxiliary wave function with an indexing vector containing a negative element are always zero. The linear HOPS equation maintains the normalization of the system reduced density matrix within the ensemble average, but the physical wave function is not normalized in individual trajectories. Instead, for long trajectories, most realizations have $\vert \vert \psi^{(\Vec{0})}\vert \vert \rightarrow 0 $ and an infinitesimal subset have physical wave functions with diverging norms.\cite{suessHierarchyStochasticPure2014} As a result, linear HOPS calculations show slow convergence with the size of the ensemble. 

We can improve convergence with ensemble size by using the non-linear HOPS equation which describes the time evolution of a normalizable stochastic wave function. We can rewrite the reduced system density matrix (eq. \eqref{eq:reduced_den_linear}) in terms of a normalized wave function and the norm contribution
\begin{flalign}
\begin{aligned}
\label{eq:reduced_den_nonlinear}
    \rho_S &= \mathbb{E}[ \vert \vert \tilde{\psi}(t; z_{n,t}) \vert \vert ^2 \vert \psi(t; z_{n,t}) \rangle \langle \psi(t; z_{n,t}) \vert]\\ &= \tilde{\mathbb{E}}[ \vert \psi(t; z_{n,t}) \rangle \langle \psi(t; z_{n,t}) \vert].
\end{aligned}
\end{flalign}
The norm in the first expression can be interpreted as a weighting factor for a new ensemble average. Using a Girsanov transform, we can solve for the corresponding equation-of-motion \cite{diosiNonMarkovianQuantumState}, which gives the non-linear HOPS equation \cite{suessHierarchyStochasticPure2014}  
\begin{flalign}
\begin{aligned}
\label{eq:NonLinearHops}
\partial_t \vert \psi^{(\Vec{k})}(t; z_{n,t}) \rangle 
=  &\big(-i\hat{H}_S - \Vec{k} \cdot \Vec{\gamma} + \sum_{n} \hat{L}_{n} (z^*_{n,t}+ \xi_{n,t})\big)\vert \psi^{(\Vec{k})}(t; z_{n,t}) \rangle \\ 
+ &\sum_{n} \Vec{k}[n] g_{n} \hat{L}_{n}  \vert \psi^{(\Vec{k} -\Vec{e}_{n})}(t; z_{n,t}) \rangle \\
- &\sum_{n} (\hat{L}^{\dagger}_{n} - \langle\hat{L}^{\dagger}_{n}\rangle_{t}) \vert \psi^{(\Vec{k}+\Vec{e}_{n})}(t; z_{n,t})\rangle, 
\end{aligned}
\end{flalign}
where
\begin{equation}
    \xi_{t,n} = \int_{0}^{t} ds \alpha^{*}_{n}(t-s) \braket{L^{\dagger}_{n}}_s
\end{equation}
is a memory term that causes a drift in the effective noise, and
\begin{equation}
    \langle\hat{L}^{\dagger}_{n}\rangle_{t} = \frac{\langle \psi^{(\vec{0})}(t; z_{n,t}) \vert \hat{L}^{\dagger}_{n}\vert \psi^{(\vec{0})}(t; z_{n,t}) \rangle}{\braket{\psi^{(\vec{0})}(t; z_{n,t}) \vert \psi^{(\vec{0})}(t; z_{n,t})}}.
\end{equation}
We note that non-linear HOPS equation does not actually normalize the physical wave function, but ensures that the contribution of each wave function is normalized in the reduced density matrix. In the following, we will drop the explicit $z_{n,t}$ dependence from the wave function for simplicity $\big(\vert \psi^{(\Vec{k})}(t; z_{n,t}) \rangle \rightarrow \vert \psi^{(\Vec{k})}_t \rangle\big)$.

The HOPS equations are a numerically convenient, formally exact expression for exciton dynamics in small molecular aggregates. Moreover, the calculations are `embarrassingly' (also called `perfectly') parallel \cite{herlihy2020art} due to the independence of individual trajectories, and, as a result, HOPS ensembles can be computed using thousands of CPUs simultaneously without loss of efficiency. The application of HOPS to large molecular aggregates, however, is limited by the scaling of the HOPS basis with the number of molecules. It is convenient to think of HOPS calculations as depending on two basis sets: the state basis ($\mathbb{S}$) and the auxiliary basis ($\mathbb{A}$). The complete state basis is a finite set of vectors that span the Hilbert space of the system, while the complete auxiliary basis is composed of an infinite set of auxiliary wave functions indexed by vectors $\Vec{k}$. To construct a finite auxiliary basis the infinite hierarchy must be truncated. Here, we employ the common triangular truncation condition which limits the auxiliary basis to those wave functions with index vectors ($\Vec{k}$) with a sum of elements less than a preselected bound $k_{max}$ ($\{\Vec{k} \in \mathbb{A}:  \sum_i k[i] \leq k_{max}\}$). If we assume one independent environment per state, then the number of auxiliary wave functions included in the triangular truncation scales as ${N_{state}+k_{max} \choose k_{max}}$ which gives an overall $\mathcal{O}(N_{state}^{k_{max}})$ scaling for large aggregates. While convergence as a function of $k_{max}$ is guaranteed, the requisite number of auxiliary wave functions is often prohibitive. 

\subsection{Short-time correction and Markovian modes}
The Drude-Lorentz correlation function given in eq. \eqref{eq:alpha_dl_coarse} has a discontinuity at t=0 arising from the symmetry condition
\begin{equation}
    \alpha(t) = \alpha^*(-t),
\end{equation}
and can be more completely written as\cite{may2008charge}
\begin{equation}
    \alpha_{dl}(t) = \big(\textrm{Re}[g_n] + i\textrm{sgn}(t)\textrm{Im}[g_n]\big)e^{-\gamma_n \vert t \vert/\hbar}.
\end{equation}
The discontinuity in the correlation function introduces a numerically inconvenient infinitely high-frequency component to the stochastic noise trajectories $z_{n,t}$. 

We ameliorate this problem by redefining the positive-time correlation function in terms of two continuous exponential functions $\alpha(t) = \alpha_{0}(t) + \alpha_{mark}(t)$ where 
\begin{equation}
    \alpha_0(t) = g_n e^{-\gamma_n \vert t \vert/\hbar}
\end{equation}
and 
\begin{equation}
    \alpha_{mark}(t) = -i\textrm{Im}[g]e^{-\gamma_{mark} \vert t \vert/\hbar}.
\end{equation}
The definition of $\alpha_{mark}(t)$ ensures the imaginary component of the total correlation is 0 when t=0, and it also provides for a smooth transition back to the naive correlation function $\alpha_0(t)$ on a finite timescale given by $\hbar/\gamma_{mark}$. We select $\gamma_{mark} = 500 \textrm{ cm}^{-1}$ for all calculations presented in the main text (except where noted) because this was sufficiently fast to ensure the the precise timescale had no influence on the calculated dynamics. 

Due to the extremely rapid timescale on which $\alpha_{mark}(t)$ decays, this mode is Markovian and high-lying contributions to the hierarchy can be neglected. In the following, we will only include the first order terms associated with these Markovian modes and we will neglect these terms in our discussion of the auxiliary wave functions forming the hierarchy. This can be viewed as a smoothing of  the noise trajectories ($z_{n,t}$) on timescales fast compared to all other dynamics.

This problem can be avoided entirely by using a different spectral density which more naturally accounts for the short-time imaginary component of the correlation function, for example, the recently reported alternative to the Drude-Lorentz oscillator with improved low-temperature behavior \cite{ishizakiPrerequisitesRelevantSpectral2020}). 

\section{Adaptive HOPS (adHOPS)}

Within the quantum state diffusion formalism, stochastic wave functions localize in the presence of thermal environments. \cite{schackQuantumStateDiffusion1995, gisinQuantumStateDiffusion1993, gisinQuantumStateDiffusion1993a} Previously, Markovian quantum state diffusion calculations have leveraged the locality of the exciton to reduce computational complexity using both a moving basis \cite{schackQuantumStateDiffusion1995, schackQuantumstateDiffusionMoving1996} and an adaptive basis. \cite{gaoChargeEnergyTransfer2019} Both of these approaches, however, require the conservation of probability which is violated in the HOPS equations because amplitude in the auxiliary wave functions can be created and destroyed. 

Here, we develop an adaptive solution to the HOPS equation-of-motion that achieves size-invariant computational scaling (i.e. $\mathcal{O}(1)$ scaling) for calculations of large molecular aggregates. We first establish a normalized non-linear HOPS equation which ensures that the magnitude of derivative terms does not diverge with increasing depth of the hierarchy. We then demonstrate how locality appears within the hierarchy of auxiliary wave functions, with a particular emphasis on the connection between locality and the flux between neighboring auxiliary wave functions. Finally, we present an adaptive algorithm for the normalized non-linear HOPS equation that satisfies a user-selected bound on the absolute derivative error.  

\subsection{Normalization of HOPS}
To ensure that the magnitude of the derivative elements for auxiliary wave functions have a consistent absolute scale across the hierarchy, we: (1) enforce normalization of the physical wave function in the time-evolution equation, and (2) redefine the auxiliary wave function coefficients. 

To enforce the normalization of the physical wave function, we rewrite the non-linear HOPS equation in terms of a normalized physical wave function. Starting with eq. \eqref{eq:NonLinearHops}, dividing all wave functions by the norm of the physical wave function, taking the derivative, and expanding terms gives 
\begin{equation} 
\begin{split}
\partial_t \vert \psi^{(\Vec{k})}_{t} \rangle = \big(-i\hat{H} - \Vec{k} \cdot \Vec{\gamma} -\Gamma_t + \sum_{n} \hat{L}_{n} (z^*_{n,t}+ \xi_{n,t})\big) \vert \psi^{(\Vec{k})}_{t} \rangle
\\ + \sum_{n} \Vec{k}[n]g_n \hat{L}_{n} \vert \psi^{(\Vec{k} -\Vec{e}_{n})}_{t}\rangle - \sum_{n} (\hat{L}^{\dagger}_{n} - \braket{\hat{L}^{\dagger}_n}_{t}) \vert \psi^{(\Vec{k}+\Vec{e}_{n})}_{t} \rangle,
\end{split}
\end{equation}
where 
\begin{flalign}
\begin{aligned}
    \Gamma_t = &\sum_{n} \braket{\hat{L}_{n}}_{t} \textrm{Re}[z^*_{n,t}+ \xi_{n,t}]
    - \sum_{n} \textrm{Re}[\braket{\psi^{(\Vec{0})}_{t} |L^{\dagger}_{N}| \psi^{(\Vec{e}_{n})}_{t}}] \\
    + &\sum_{n} \braket{\hat{L}^{\dagger}_n}_{t} \textrm{Re}[\braket{\psi^{(\Vec{0})}_{t} | \psi^{(\Vec{e}_{n})}_{t}}] 
\end{aligned}
\end{flalign}
is the normalization correction factor. 

\begin{figure}
	\centering
	\includegraphics[width=\linewidth]{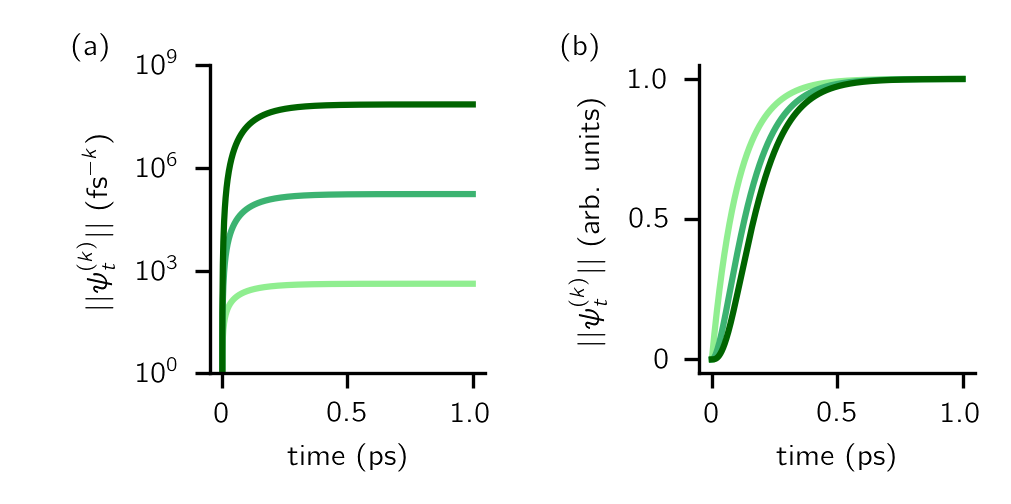}
	\caption{Magnitude of the first three auxiliaries in a single-mode HOPS calculation. \textbf{(a)} The magnitude of the auxiliaries calculated using the non-linear HOPS equation (darker lines correspond to higher k values). \textbf{(b)} The magnitude of the auxiliaries calculated using the non-linear HOPS equation with the k-dependent prefactor. Parameters: $\lambda$ = 50 cm$^{-1}$, $\gamma = 50$ cm$^{-1}$, T = 295 K, and $k_{max} = 10$. No Markovian mode was included in this calculation.}
	\label{fig:aux_norm}
\end{figure}

In the non-linear HOPS equation, the magnitude of the auxiliary wave functions grows with increasing auxiliary index. The basic HOPS terminator \cite{suessHierarchyStochasticPure2014} for a hierarchy with a single thermal environment is 
\begin{equation}
    \psi^{(k)}_{t} = \frac{g}{\gamma}\hat{L}\psi^{(k-1)}_{t}.
\end{equation}
When this terminator is used for the first-order auxiliaries the resulting equation is equivalent to the standard Markovian quantum state diffusion equation. Fig. \ref{fig:aux_norm}a, shows the norm of the first three auxiliary wave functions for a single trajectory with a hierarchy consisting of one Drude-Lorentz oscillator. The magnitude $\vert \vert \psi^{(\Vec{k})}_t\vert\vert$ increases with increasing auxiliary index which, given $g/ \gamma>1$, is consistent with the terminator condition. For a single mode hierarchy, we can ensure the norm of the auxiliary wave functions does not diverge by introducing a new k-dependent prefactor for each wave function $(\gamma/g)^{k}$, as shown in Figure \ref{fig:aux_norm}(b). In the multimode case, we extend the definition of the prefactor to $\prod_{n}(\gamma_{n}/g_{n})^{\Vec{k}[n]}$ which ensures that the auxiliary wave functions that define the edges of the hierarchy (only one non-zero mode) do not diverge with increasing hierarchy depth. Rewriting the non-linear HOPS equation to account for this additional prefactor leads to the normalized non-linear HOPS equation
\begin{flalign}
\begin{aligned}
\label{eq:NormNonLinearHops}
\partial_t \vert \psi^{(\Vec{k})}_t \rangle 
=  \big(-i\hat{H}_S - \Vec{k} \cdot \Vec{\gamma} -\Gamma_t + \sum_{n} \hat{L}_{n} (z^*_{n,t}+ \xi_{n,t})\big)&\vert \psi^{(\Vec{k})}_t \rangle \\ 
+ \sum_{n} \Vec{k}[n] \gamma_{n} \hat{L}_{n}  &\vert \psi^{(\Vec{k} -\Vec{e}_{n})}_t \rangle \\
- \sum_{n} (\frac{g_n}{\gamma_{n}})(\hat{L}^{\dagger}_{n} - \langle\hat{L}^{\dagger}_{n}\rangle_{t}) &\vert \psi^{(\Vec{k}+\Vec{e}_{n})}_t\rangle, 
\end{aligned}
\end{flalign}
where
\begin{flalign}
\begin{aligned}
    \Gamma_t = \sum_{n} \braket{\hat{L}_{n}}_{t} \textrm{Re}[z^*_{n,t}+ \xi_{n,t}]
    &- \sum_{n} (\frac{g_n}{\gamma_{n}}) \textrm{Re}[\braket{\psi^{(\vec{0})}_t |\hat{L}^{\dagger}_{n}| \psi^{(\vec{e}_{n})}_t}] \\
    &+ \sum_{n} (\frac{g_n}{\gamma_{n}}) \braket{\hat{L}^{\dagger}_n}_{t} \textrm{Re}[\braket{\psi^{(\Vec{0})}_t | \psi^{(\Vec{e}_{n})}_t }]
\end{aligned}
\end{flalign}
ensures normalization of the physical wave function. 

\subsection{Locality of HOPS}
To construct an adaptive approach to solving the HOPS equations, we must first address the central question: How and to what extent does the locality expected in the quantum state diffusion formalism appear in HOPS? 

\begin{figure}
	\centering
	\includegraphics{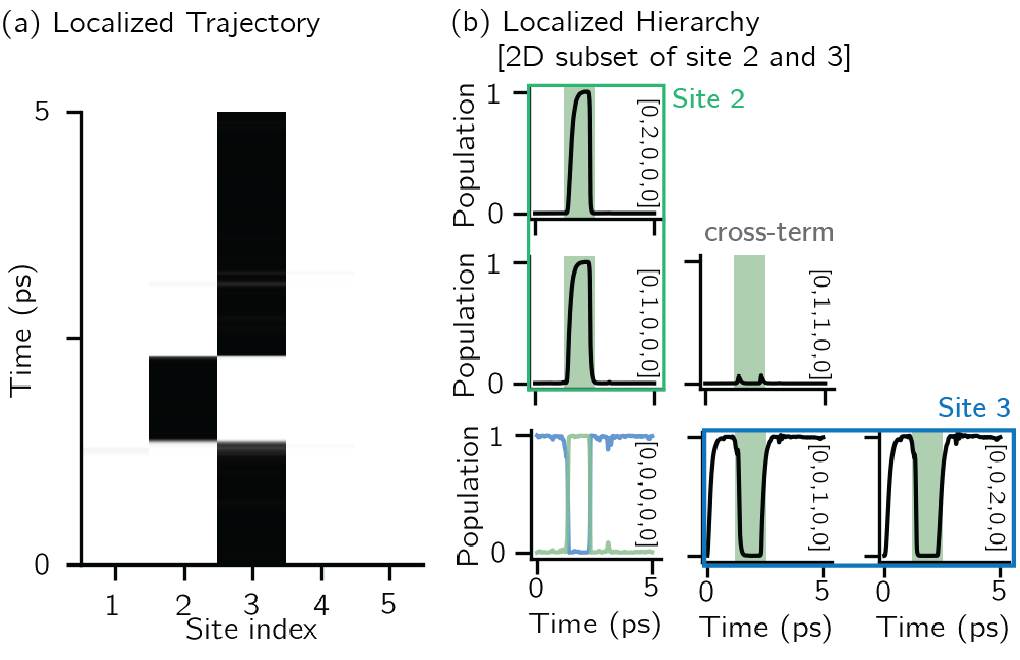}
	\caption{Localization in a single HOPS trajectory.  \textbf{(a)} Contour map of site populations (darker is more populated). \textbf{(b)} Norm-squared of auxiliary wave functions for a two-dimensional subset of the hierarchy associated with site 2 (column) and 3 (row). Panels are labelled by their index vector ($\Vec{k}$). The shaded region represents the time-period when site 2 is occupied. The physical wave function ($\Vec{k} = \Vec{0}$) shows the populations of site 2 and 3 as green and blue lines, respectively. Parameters: V = 10 cm$^{-1}$, $\lambda$ = 50 cm$^{-1}$, $\gamma = 50$ cm$^{-1}$, T = 295 K, and $k_{max} = 10$.}
	\label{fig:locality}
\end{figure}

Fig. \ref{fig:locality} shows that in HOPS calculations, localization in the physical wave function induces localization in the hierarchy. By `localization in the hierarchy,' we specifically refer to clustering of amplitude in a small set of auxiliary wave functions in a way that depends on the position of the excitation in the physical wave function. In Fig. \ref{fig:locality}a, an excitation begins on the middle site of a five pigment chain. The coupling to the thermal environment induces substantial localization and the excitation jumps between site 3 and site 2. Fig. \ref{fig:locality}b plots the norm of auxiliary wave functions associated with site 2 and site 3; each plot is labeled by an auxiliary vector index $\Vec{k}$. For example, [0,1,0,0,0] (Fig. \ref{fig:locality}b, first column) is the first order auxiliary wave function associated with the thermal environment on the second site. The occupation of the auxiliary wave functions (black lines, Fig. \ref{fig:locality}b) track with the population of the true wave function - i.e., the auxiliary wave functions associated with the thermal environment of site 2 (first column Fig. \ref{fig:locality}b) are only occupied when site 2 is occupied in the true wave function (shaded region).

The locality in the HOPS hierarchy can be understood in terms of the balance of flux terms in the normalized non-linear HOPS equation. First, every auxiliary wave function is damped (first line of eq. \eqref{eq:NormNonLinearHops}, $-(\Vec{k}\cdot \Vec{\gamma}) \vert \psi^{(\Vec{k})}_t\rangle$ ) and therefore has zero amplitude without a continuous source term. The fundamental source term is the physical wave function which is the lowest order of the hierarchy. The flux of amplitude towards higher-lying auxiliary wave functions arises from the second line of eq. \eqref{eq:NormNonLinearHops} ($\Vec{k}[n] \gamma_{n} \hat{L}_{n}  \vert \psi^{(\Vec{k} -\Vec{e}_{n})}_t \rangle$). For system-bath coupling operators that are site projection operators ($\hat{L}_n = \vert n \rangle \langle n \vert$), the flux towards higher-lying auxiliaries will only arise when there is amplitude on the associated site of the lower auxiliary wave function. Moreover, the auxiliary wave functions are localized by the same dynamics that localize the physical wave function (first line, eq. \eqref{eq:NormNonLinearHops}). As a result, the localized auxiliary wave functions will only contribute amplitude to higher-lying auxiliary wave functions with an index that differs by $+\Vec{e}_n$ in a site (n) with non-zero amplitude. Thus, the locality of the physical wave function results in preferential population of the specific auxiliary wave functions. 

\subsection{Adaptive algorithm}
We have developed an adaptive algorithm for time-evolving the HOPS equations (adaptive HOPS, adHOPS) that leverages locality by constructing a reduced basis set at each time point that is still capable of describing the full dynamics. We establish the essential basis set elements at each time point (t) by ensuring that the error in the time-derivative introduced by the truncated auxiliary ($\mathbb{A}_t$) and state ($\mathbb{S}_t$) basis is below a given threshold ($\delta$). We define the derivative error in terms of Euclidean distance between the true derivative vector and the effective derivative vector constructed using the adaptive basis. The key equations (see below) provide an upper bound on the derivative error squared and are derived by considering all possible flux contributions in the normalized non-linear HOPS equation (eq. \eqref{eq:NormNonLinearHops}), excluding higher order effects introduced through the normalization correction ($\Gamma_t$). Because auxiliary wave functions share only nearest neighbor connections ($\Vec{k}\pm\Vec{e}_n \leftarrow \Vec{k}$) and the Hamiltonian for a molecular aggregate supports electronic couplings over a finite spatial extent, the adaptive basis can be constructed with $\mathcal{O}(1)$ scaling. The result is a calculation where in addition to a trajectory of the wave functions, we construct an adaptive basis-set trajectory $\{\mathbb{B}_{0}, \mathbb{B}_{dt}, ..., \mathbb{B}_t\}$. 

Our adHOPS algorithm neither assumes nor imposes locality. Rather, the adaptive basis takes advantage of whatever locality arises during the simulation. If the full hierarchy is required to satisfy the derivative error bound, adHOPS smoothly reverts to a HOPS calculation. As a result, adHOPS remains formally exact - the adaptive basis represents a time-dependent truncation of hierarchy elements, and $\delta$, like $k_{max}$, is a convergence parameter.

We note that the current adHOPS algorithm makes use of two approximations: first, the spectral density is assumed to be over-damped (e.g. Drude-Lorentz), which allows for a consistent normalization of the hierarchy elements. Second, the system-bath coupling operator is assumed to be a site-projection operator ($L_n = \vert n \rangle \langle n \vert$); in other words, we assume that each molecule has an independent vibrational environment.

\underline{Basis Sets.}

The adaptive basis at the previous time point (t) is defined as the direct sum of a truncated auxiliary and state basis ($\mathbb{B}_t = \mathbb{A}_{t} \bigoplus \mathbb{S}_{t}$). In practice, this means that when a state n is not in the adaptive state basis at the previous time point ($n \notin \mathbb{S}_t$) the coefficient of that state is necessarily zero for all auxiliary wave functions at that time point (and vice-versa for an auxiliary $\vec{k} \notin \mathbb{A}_t$). In the following, we will refer to auxiliaries ($\Vec{k} \in \mathbb{A}_t$) and states ($n \in \mathbb{S}_t$) which belong to the adaptive basis at the previous time point as `populated' because they are the only elements with non-zero coefficients. 


Given an adaptive basis at the previous time point, the challenge of constructing the new adaptive basis can be split into two pieces: constructing the new auxiliary basis ($\mathbb{A}_{\textrm{new}}$) and the new state basis ($\mathbb{S}_{\textrm{new}}$). For each of these pieces, we will need to determine which populated elements will remain in the new basis ($\mathbb{A}_p, \mathbb{S}_p$) and which elements that were not in the previous basis will need to be included, what we will refer to as the boundary elements ($\mathbb{A}_b, \mathbb{S}_b$). The new adaptive basis is $\mathbb{B}_{t+dt} = \mathbb{A}_{\textrm{new}} \bigoplus \mathbb{S}_{\textrm{new}}$, where $\mathbb{A}_{\textrm{new}} = \mathbb{A}_p \cup \mathbb{A}_b$ and $\mathbb{S}_{\textrm{new}} = \mathbb{S}_p \cup \mathbb{S}_b$. At each step, we ensure that the error introduced by truncating the basis set is bounded below an error threshold $\delta_{A/S,p/b}$ such that $\delta^2 = \delta_{A,p}^2 + \delta_{A,b}^2 + \delta_{S,p}^2 + \delta_{S,b}^2$. Because the total basis is constructed as the direct sum of the auxiliary and state bases, there is a coupling between the construction of the new auxiliary basis and the new state basis. As a result, the order of basis set construction can influence the results. Though these operations can be performed in any order, in the equations below we assume that the new auxiliary basis is constructed before the new state basis. 

For clarity, we assume there is a single thermal environment associated with each molecule. The equations we provide have straightforward extensions for the case of multiple thermal environments on each molecule.   

\underline{Flux components.}

In the following we will consider how different flux components contribute to the error when basis set elements are neglected in the adaptive basis. To simplify that presentation, we begin by decomposing the components of eq. \eqref{eq:NormNonLinearHops} into convenient pieces. In the following we revert to a vector notation for the auxiliary wave functions, so $\psi_t^{(\vec{k})}[n]$ is the coefficient of the $n$ state on the $\Vec{k}$ auxiliary wave function. 

It will be convenient to group fluxes that connect populated basis set elements
\begin{equation}
    D^{(\Vec{k},\Vec{k}')} [n, n'] = \hat{K}_{\Vec{k}, n \leftarrow \Vec{k}', n'}(t) \psi_t^{(\vec{k}')}[n']
\end{equation}
where $\hat{K}_{\Vec{k}, n \leftarrow \Vec{k}', n'}(t)$ is the time-evolution operator for the complete hierarchy constructed using the adaptive basis for the previous time point ($\mathbb{B}_t$). We will make use of the fact that 
\begin{equation}
\label{eq:deriv_sum_property}
    \partial_t \psi_t^{(\vec{k})}[n] = \sum_{(\Vec{k}', n')\in \mathbb{B}_t} D^{(\Vec{k},\Vec{k}')} [n, n']
\end{equation}
as long as $(\vec{k}, n) \in \mathbb{B}_t$. 

We will also need to decompose the different flux contributions into three basic groups. First, those fluxes that can change the state index 
\begin{equation}
    F^{(\vec{k})}[m,n] = -i\hat{H}_S[m,n] \psi_t^{(\vec{k})}[n]
\end{equation}
which arise from system Hamiltonian inducing coupling between states within a single auxiliary, on the first line of eq. \eqref{eq:NormNonLinearHops}. Second, those fluxes that can increase the auxiliary index ($\Vec{k}+\Vec{e}_n \leftarrow \vec{k}$)
\begin{equation}
    I^{\Vec{k}}_{+}[n] = \gamma_n (\Vec{k}[n]+1) \psi_t^{(\vec{k})}[n]
\end{equation}
which arises from the second line of eq. \eqref{eq:NormNonLinearHops}. Third, those fluxes that can decrease the auxiliary index ($\Vec{k}- \Vec{e}_n \leftarrow \vec{k}$) which we divide into
\begin{equation}
    I^{\Vec{k}}_{-}[n] =  -\frac{g_n}{\gamma_n} \psi_t^{(\vec{k})}[n]
\end{equation}
arising from the $\hat{L}_n$ in the third line of eq. \eqref{eq:NormNonLinearHops} and 
\begin{equation}
    G^{\Vec{k}}_{-,n}[m] = \frac{g_n}{\gamma_n} \braket{\hat{L}_n}_t \psi_t^{(\vec{k})}[m]
\end{equation}
where the $\braket{\hat{L}_n}_t$ term allows flux from the $m$ state coefficients on the $\vec{k}$ auxiliary to the $m$ state coefficients on the $\Vec{k}-\Vec{e}_n$ auxiliary.




\underline{Auxiliary Basis: Populated Wave Functions.}

The first step in constructing the new auxiliary basis is to determine which of the populated auxiliary wave functions ($\Vec{k} \in \mathbb{A}_t$) can be neglected while ensuring the associated derivative error is below the threshold $\delta_{A,p}$. To determine the error associated with neglecting one populated auxiliary wave function, we consider all of its possible contributions to the derivative vector. 

The simplest contribution is the derivative of the coefficients for each populated state in the auxiliary vector. Using the sum property defined in eq. \eqref{eq:deriv_sum_property} we can write this squared error term as 
\begin{flalign}
\begin{aligned}
\label{eq:err_aux_pop_flux_in}
&\sum_{n \in \mathbb{S}_t} \vert \sum_{(\Vec{k}', n')\in \mathbb{B}_t} D^{(\Vec{k},\Vec{k}')} [n, n']\vert^2\\
&=\vert \vert \partial_{t, \mathbb{B}_t}\vert \psi^{(\Vec{k})}_t \rangle \vert \vert^2
\end{aligned}
\end{flalign}
where in the second line we have written $\partial_{t, \mathbb{B}_t}$ to remind us that this equation is an abridgement that only holds for the populated states of the $\Vec{k}$ auxiliary wave function (i.e. the components of the auxiliary wave function that are in the adaptive basis at the previous time point). 

In addition to their own derivative components, the populated auxiliary $\Vec{k}$ can also contribute to the derivative by providing flux. We avoid double counting error terms included in eq. \eqref{eq:err_aux_pop_flux_in}, by only considering non-populated states ($m \notin \mathbb{S}_t)$ in the squared error term associated with the $\Vec{k} \leftarrow \Vec{k}$ flux 
\begin{flalign}
\begin{aligned}
\label{eq:err_aux_pop_flux_0}
    &\sum_{m\notin \mathbb{S}_t}  \vert \sum_{n \in \mathbb{S}_t} F^{(\Vec{k})}[m,n]  \vert^2\\
    & = \vert \vert (\hat{H} - \hat{P}_{\mathbb{S}_t} \hat{H} \hat{P}_{\mathbb{S}_t}) \vert \psi^{(\Vec{k})}_t\rangle \vert \vert^2
\end{aligned}
\end{flalign}
where the second line is a convenient operator expression for these terms making use of $\hat{P}_{\mathbb{S}_t}$, the operator that projects onto the populated states ($\mathbb{S}_t$). The squared error arising from the flux towards auxiliaries with a larger index is given by 
\begin{equation}
\label{eq:err_aux_pop_flux_-}
    \sum_{n \in \mathbb{S}_t} \vert I_{+}^{\Vec{k}}[n]\vert^2 \Theta[\Vec{k} + \Vec{e}_n, \mathbb{A}]
\end{equation}
where  
\begin{equation}
\Theta[\vec{k}, \mathbb{A}] = 
\begin{cases} 
1 &\textrm{ if } \Vec{k} \in \mathbb{A} \\
0 &\textrm{ otherwise}
\end{cases} 
\end{equation}
ensures we only consider flux terms that lead to legal members of the auxiliary basis (since all others should be neglected). The squared error associated with the flux towards auxiliaries with a smaller index is given by
\begin{flalign}
\begin{aligned}
\label{eq:err_aux_pop_flux_+}
    &\sum_{n \in \mathbb{S}_t} ( \vert I_{-}^{\Vec{k}}[n] + G_{-,n}^{\Vec{k}}[n] \vert^2 + \sum_{n \neq m\in \mathbb{S}_t}  \vert G_{-,n}^{\Vec{k}}[m] \vert^2)\Theta[\Vec{k} - \Vec{e}_n, \mathbb{A}]\\
    &\leq \sum_{n \in \mathbb{S}_t} \Theta[\Vec{k} - \Vec{e}_n, \mathbb{A}] \Big\vert \frac{g_n}{\gamma_n} \Big\vert^2 (\vert \psi^{(\Vec{k})}_t[n]\vert^2 + \langle\hat{L}^{\dagger}_{n}\rangle_{t}^2 \vert \vert \psi^{(\Vec{k})}_t \vert \vert^2)
\end{aligned}
\end{flalign}
where we have rearranged terms in the second line to generate a more convenient expression, at the price of introducing an upper bound. 

If we neglect the $\Vec{k}$ auxiliary wave function, then by the end of the next time step the coefficients are forced to zero. This implicitly introduces a fictitious derivative constructed to precisely cancel the current amplitude in a single time step. Since this flux does not arise in the HOPS equation, this is an additional squared error term in our derivative
\begin{equation}
\label{eq:error_aux_pop_implicit}
    \vert \vert \psi^{(\Vec{k})}_t\vert \vert^2/\Delta t^2
\end{equation}
which depends on the simulation time step ($\Delta t$). 

The square of the derivative error introduced by removing the $\Vec{k}$ populated auxiliary wave function ($E_p^2[\Vec{k}]$) is bounded by the sum of eqs. \eqref{eq:err_aux_pop_flux_in}, \eqref{eq:err_aux_pop_flux_0}, \eqref{eq:err_aux_pop_flux_-}, \eqref{eq:err_aux_pop_flux_+}, and \eqref{eq:error_aux_pop_implicit}. We determine the largest set of auxiliaries that can be removed at the current time point while maintaining the bound $\delta_{A,p}$ on the derivative error. The remaining auxiliaries, those that were in the adaptive basis in the previous time point and will be in the adaptive basis in the next time point, define the set $\mathbb{A}_p$. We note that our selection criterion (maximum number of auxiliaries removed), like all subsequent basis set selections, is not unique and a variety of different algorithms can be used to determine which auxiliaries to keep at each time point while satisfying the error bound.

\underline{Auxiliary Basis: Boundary Wave Functions.}

Auxiliary wave functions that are members of the full auxiliary basis but were not in the adaptive basis at the previous time point ($\Vec{k} \in \mathbb{A}\setminus\mathbb{A}_t$) have no amplitude to contribute to flux but may still be important to the overall dynamics by accepting amplitude from populated auxiliaries. Naively, one might attempt to calculate the error for neglecting each possible boundary auxiliary $\Vec{k} \in \mathbb{A}\setminus\mathbb{A}_t$ which would scale with the size of the full auxiliary basis and be unmanageable for even moderately sized pigment aggregates. However, the only way for an auxiliary $\Vec{k}$ to belong to this set is for it to be connected to one (or more) populated auxiliaries. As a result, it is more efficient to determine the important connections with populated auxiliaries than to directly search for the important boundary auxiliary wave functions. 

We can determine an upper bound on the squared error for neglecting boundary auxiliary wave functions in terms of the populated auxiliary $\Vec{k}' \in \mathbb{A}_p$ that creates the flux and the mode (n) along which it is connected to the boundary, either from below $(\Vec{k}', n, +)$:
\begin{flalign}
\begin{aligned}
   \big(\Theta[\vec{k}' + \vec{e}_n, \mathbb{A}\setminus\mathbb{A}_t] \vert I_{+}^{\Vec{k}'}[n]\vert^2\big),
\end{aligned}
\end{flalign}
or from above $(\Vec{k}', n, -)$:
\begin{equation}
    \Theta[\Vec{k}' - \Vec{e}_n, \mathbb{A}\setminus\mathbb{A}_t] \Big\vert \frac{g_n}{\gamma_n} \Big\vert^2 (\vert \psi^{(\Vec{k}')}_t[n]\vert^2 + \langle\hat{L}^{\dagger}_{n}\rangle_{t}^2 \vert \vert \psi^{(\Vec{k}')}_t \vert \vert^2)
\end{equation}
where the second expression arises from the same considerations leading to eq. \eqref{eq:err_aux_pop_flux_+} and is an upper bound. We introduced $\Theta[\vec{k}' \pm \vec{e}_n, \mathbb{A}\setminus\mathbb{A}_t]$ operators to ensure that the each flux term goes to an auxiliary wave function that was not in the adaptive basis at the previous time point. 

Treating each of these error terms independently, we construct the largest set of tuples that can be removed $\{(\Vec{k}', n, \pm),... \}$ such that the associated error is less than $\delta_{A,b}$. The set $\mathbb{A}_b$ is composed of all auxiliaries constructed from the remaining tuples ($\vec{k} = \vec{k}'\pm \vec{e}_n$). This algorithm does not guarantee that the minimal error is achieved since we do not determine which auxiliary each flux term leads to until after the truncation. However, it has the advantage of introducing only a small additional computational cost since the vast majority of all connections to the boundary are negligible due to localization in the auxiliary wave functions.

\underline{State Basis: Populated States.}

To strengthen the analogy between the auxiliary and state bases, we introduce a new vector $\vert \phi^{(n)}_t \rangle$ which contains the coefficient of the $n^{th}$ state across all auxiliaries in the reduced set $\mathbb{A}_p$ (i.e. $\phi^{(n)}_t[\vec{k}] = \psi^{(\Vec{k})}_t[n]$ if $\vec{k} \in\mathbb{A}_p $). The construction of the state basis is completely analogous to the auxiliary basis construction. A brief description is provided below for completeness. 

For a populated state, we first consider its contribution to the derivative of coefficients for each populated auxiliary wave function 
\begin{flalign}
\begin{aligned}
\label{eq:err_state_pop_flux_in}
&\sum_{\Vec{k} \in \mathbb{A}_p} \vert \sum_{(\Vec{k}', n')\in \mathbb{A}_p\bigoplus\mathbb{S}_t} D^{(\Vec{k},\Vec{k}')} [n, n'] \vert^2\\
&=\vert \vert \hat{P}_{\mathbb{A}_p} \partial_{t,\mathbb{B}_t} (\hat{P}_{\mathbb{A}_p} \vert \phi^{(n)}_t \rangle) \vert \vert^2
\end{aligned}
\end{flalign}
where the sums in the first line only considers auxiliary wave function that are in the truncated set of populated wave functions $\mathbb{A}_p$. In the second line, we rewrite that into a convenient operator notation again recognizing the abridged time-evolution $\partial_{t, \mathbb{B}_t}$ which must be further reduced onto the truncated set of populated auxiliary wave functions by the projection operator $\hat{P}_{\mathbb{A}_p}$. 

In addition to their own derivative components, the populates state n can also contribute to the derivative by providing flux. We avoid double counting error already included in eq. \eqref{eq:err_state_pop_flux_in} by only considering non-populated states ($m \notin \mathbb{S}_t$) for the squared error term associated with the $\Vec{k}\leftarrow \Vec{k}$ flux
\begin{flalign}
\begin{aligned}
\label{eq:err_state_pop_flux_out_0}
    \sum_{\Vec{k} \in \mathbb{A}_p} \sum_{m \notin \mathbb{S}_t} \vert F^{(\Vec{k})}[m,n] \vert^2\\
   = V[n] \textrm{ } \vert \vert \phi^{(n)}_t \vert \vert^2
\end{aligned}
\end{flalign}
where in the second line we have introduced 
\begin{equation}
    V[n] = \sum_{m \notin \mathbb{S}_t} \vert H_s[m,n]\vert^2
\end{equation}
which quantifies the total coupling of state $n$ to all states not included in the previous basis ($m\notin \mathbb{S}_t$). In addition there are the flux terms which can increase the auxiliary index
\begin{equation}
\label{eq:err_state_pop_flux_-}
    \sum_{\vec{k} \in \mathbb{A}_p} \vert I_{+}^{\Vec{k}}[n]\vert^2 \Theta[\Vec{k} + \Vec{e}_n, \mathbb{A}]
\end{equation}
or decrease the auxiliary index
\begin{equation}
\label{eq:err_state_pop_flux_+}
    \sum_{\vec{k} \in \mathbb{A}_p} \Theta[\Vec{k} - \Vec{e}_n, \mathbb{A}] \Big\vert \frac{g_n}{\gamma_n} \Big\vert^2 (\vert \psi^{(\Vec{k})}_t[n]\vert^2 + \langle\hat{L}^{\dagger}_{n}\rangle_{t}^2 \vert \vert \psi^{(\Vec{k})}_t \vert \vert^2).
\end{equation}
Again, eq. \eqref{eq:err_state_pop_flux_+} represents an upper bound on the squared error. 

Finally, the squared derivative error arising from the fictitious flux required to cancel the residual amplitude on the neglected state is given by  
\begin{equation}
\label{eq:err_state_pop_implicit}
    \vert \vert \phi^{(n)}_t \vert \vert^2/\Delta t^2.
\end{equation}

Using the bound on the squared derivative error given by eqs. \eqref{eq:err_state_pop_flux_in}-\eqref{eq:err_state_pop_implicit},  we determine the largest set of states (\{$n'$\}) which can be neglected while ensuring the total error is smaller than $\delta_{S,p}$. The set of remaining states we will label $\mathbb{S}_p$.

\underline{State Basis: Boundary States} 

The set of states that are not included in the adaptive basis at the previous time point can be important for accurately propagating the time evolution if they accept flux from one (or more) populated states. When the system-bath coupling operators ($\hat{L}_n$) are site-projection operators, the only term in the normalized non-linear HOPS equation which can change the state index is the system Hamiltonian ($\hat{H}_s$), and the corresponding squared error for neglecting the flux into a state $n \in \mathbb{S}\setminus\mathbb{S}_t$ is given by 
\begin{flalign}
\begin{aligned}
\label{eq:err_state_bound_flux}
\sum_{k \in \mathbb{A}_p} \vert \sum_{m \in \mathbb{S}_p} F^{(\Vec{k})}[n,m] \vert^2\\
= \sum_{\Vec{k} \in \mathbb{A}_p} \vert \Psi^{(\vec{k})}[n] \vert^2
\end{aligned}
\end{flalign}
where 
\begin{equation}
    \vert \Psi^{(\vec{k})}_t \rangle = (\hat{H}_s - \hat{P}_{\mathbb{S}_p} \hat{H}_s \hat{P}_{\mathbb{S}_p})\hat{P}_{\mathbb{S}_p} \vert \psi^{(\Vec{k})}_t \rangle
\end{equation}
provides a convenient operator formulation. We note that switching notation for $\Psi^{(\vec{k})}$ in analogy to $\phi^{(n)}$ defined above, would allow us to write this squared error term compactly as
\begin{equation}
    \vert \vert \Phi^{(n)}_t \vert \vert^2
\end{equation}
where we understand $\Phi^{(n)}_t[\vec{k}] = \Psi^{(\Vec{k})}_t[n]$ if $\vec{k} \in\mathbb{A}_p $ and $n \in \mathbb{S}_t$.

We determine the largest set of states ($n' \in \mathbb{S}\setminus\mathbb{S}_t$) that can be neglected while maintaining $\delta_{S,b}$ as the bound on the derivative error. The remaining states form the set $\mathbb{S}_b$. 

\underline{Hamiltonian couplings}

To achieve $\mathcal{O}(1)$ scaling in eqs. \eqref{eq:err_aux_pop_flux_0}, \eqref{eq:err_state_pop_flux_out_0} and \eqref{eq:err_state_bound_flux}, the system Hamiltonian must be sparse. We note that for a physical Hamiltonian that supports coupling over a finite spatial extent (e.g., $r ^{-3}$ scaling of dipole-dipole coupling), this sparsity requirement is necessarily fulfilled for large aggregates. The simplest computational approach to leveraging the locality of coupling is to filter the system Hamiltonian so that elements below a threshold ($\epsilon$) are set to zero before the calculation begins.  

\underline{Adaptive Parameters.}

The adaptive basis at each time point is defined by four error parameters ($\delta_{A,p}, \delta_{A,b}, \delta_{S,p}, \delta_{S,b}$). Instead of specifying each of the four parameters, we select a single parameter $\delta$ and, for all calculations presented here, require the error to be equally distributed between the auxiliary and state basis ($\delta_A^2 = \delta_S^2 = \delta^2/2$). The explicit distribution among the sub-parameters is determined at each time point. For the auxiliary basis, the error bound used for the populated auxiliaries is required to obey $\delta_{A,p} \leq \delta_{A}/\sqrt{2}$. This value is often not saturated, so at each time point we define $\delta_{A,b}^2 = \delta_A^2 - \delta_{A,p}^2$. The equivalent is done for the state basis. 

The algorithm above can be partitioned to allow for either the state or auxiliary basis to be treated adaptively while the other is statically defined. For small aggregates, in particular,  it is often convenient to not solve for an adaptive state basis since most (or even all) states will be in the adaptive basis most of the time. 

\section{Results and Discussion}
\begin{figure}
    \centering
    \includegraphics{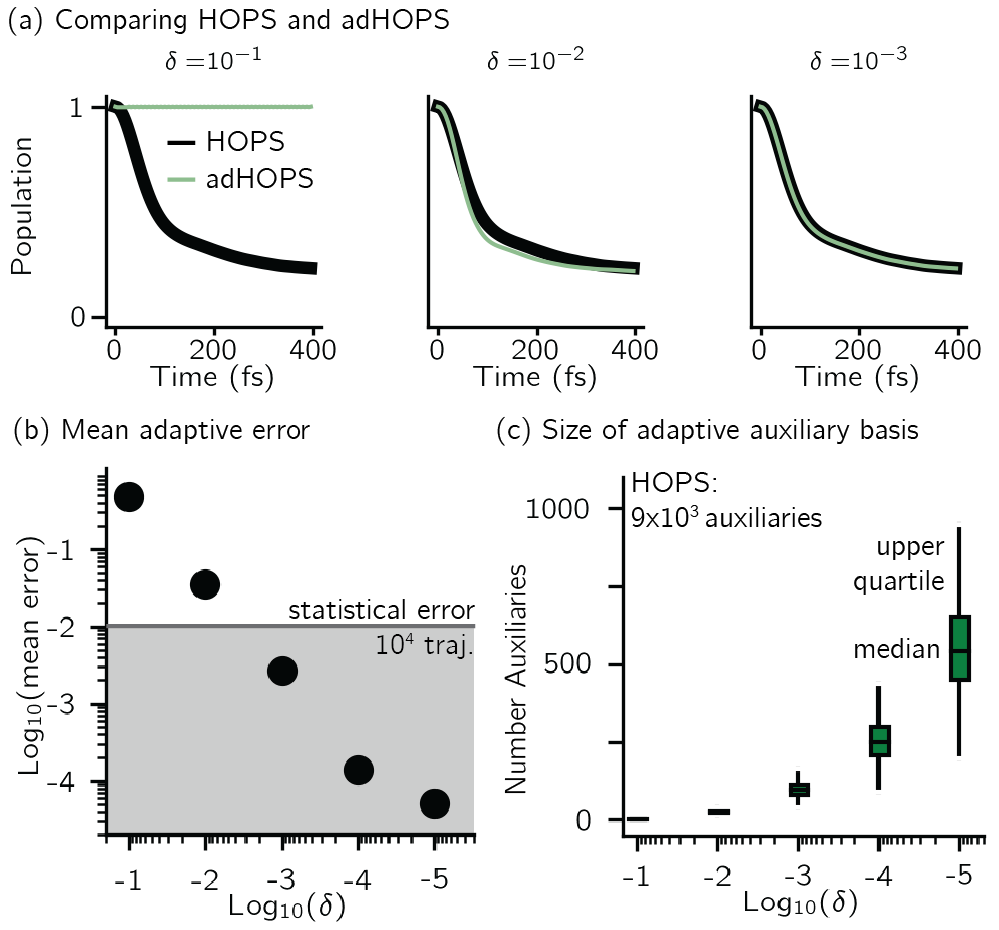}
    \caption{Comparing HOPS and adHOPS for a five-site linear chain. \textbf{(a)} Site 3 population dynamics for HOPS (black line) and adHOPS (green line). \textbf{(b)} Mean adaptive error as a function of $\delta$. The grey region represents error beneath the statistical error for a $10^{4}$ trajectory ensemble. \textbf{(c)} Ensemble distribution of the size of the adaptive auxiliary basis as a function of $\delta$. Parameters: V = 50 cm$^{-1}$, $\lambda$ = 50 cm$^{-1}$, $\gamma = 50$ cm$^{-1}$, T = 295 K, $k_{max} = 10$, and $N_{\textrm{traj}} = 10^4$.}
    \label{fig:SmallAgg_dynamics}
\end{figure}
For a five-site linear chain, adHOPS calculations converge rapidly with respect to the derivative error bound and require only a small fraction of the full HOPS basis. Fig. \ref{fig:SmallAgg_dynamics}a shows the comparison between full (black line) and adaptive (green line) HOPS population dynamics of the initially excited pigment (site 3). For $\delta=10^{-1}$, the adaptive basis set is so small that the calculation shows no excitation transport. Smaller values of $\delta$ improve the description, and by $\delta=10^{-3}$ the mean error is less than $10^{-2}$. Fig. \ref{fig:SmallAgg_dynamics}b shows the mean adaptive error as a function of $\delta$. In the grey region the adaptive error is smaller than the statistical error associated with the $10^4$ trajectory ensemble. We measure the size of the auxiliary basis for a single trajectory by the average number of auxiliary wave functions required across time points. Fig. \ref{fig:SmallAgg_dynamics}c plots the ensemble distribution of the auxiliary basis size as a function of $\delta$. For $\delta = 10^{-3}$, most adHOPS trajectories require $10^2$ auxiliaries on average, or approximately 1\% of the $9\times10^3$ auxiliaries required for a HOPS calculation. Improving the accuracy of the calculation by decreasing $\delta$ two orders of magnitude only requires about four times as many auxiliaries. The other kinds of error that arise in HOPS simulations, including statistical error from a finite number of trajectories and hierarchy error from the finite $k_{max}$ value, are reported in the Supplementary Information.$^\dag$ 

\begin{figure}
	\centering
	\includegraphics{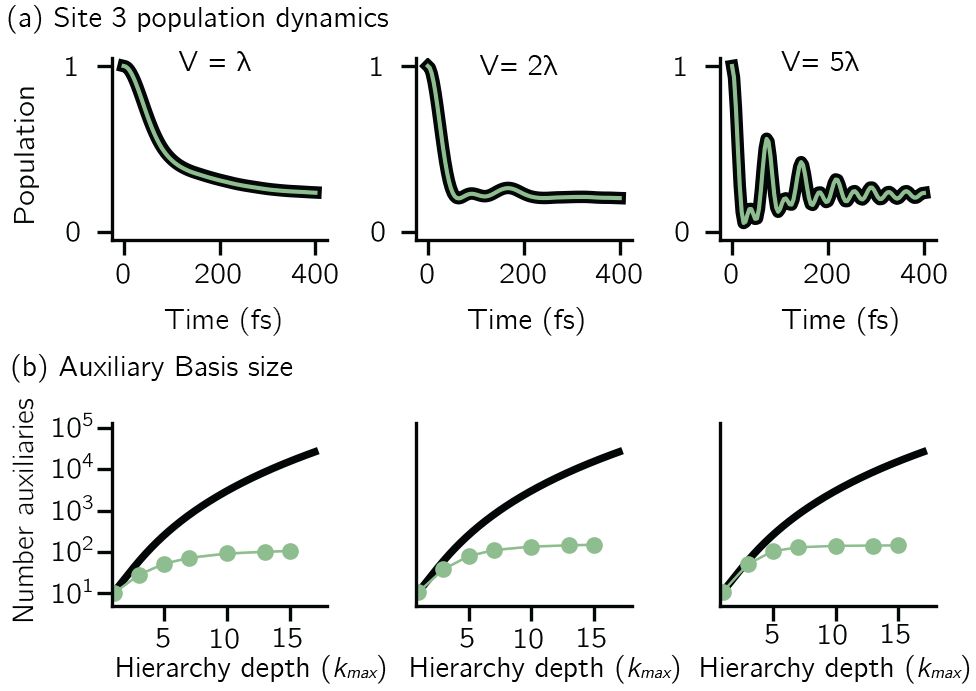}
	\caption{Comparing dynamics and auxiliary basis size as a function electronic coupling (V) for the full (black) and adaptive (green) HOPS calculations. \textbf{(a)} Site 3 population dynamics when $k_{max} = 10$. \textbf{(b)} Size of the auxiliary basis as a function of maximum hierarchy depth ($k_{max}$). Other parameters: $\lambda$ = 50 cm$^{-1}$, $\gamma = 50$ cm$^{-1}$, $T$ = 295 K, $\delta = 10^{-3}$, and $N_{\textrm{traj}} = 10^4$. For V= $250 \textrm{ cm}^{-1}$, $\gamma_{mark} = 1000 \textrm{ cm}^{-1}$, all others used $\gamma_{mark} = 500 \textrm{ cm}^{-1}$.} 
	\label{fig:adHOPS_Vscan}
\end{figure}

One persistent challenge for numerical implementations of formally exact methods is demonstrating the calculations are converged to the exact answer. In hierarchical methods, calculations must be converged with respect to the auxiliary basis which is defined in the triangular truncation condition by the maximum hierarchy level considered ($k_{max}$). In HOPS, the criterion for convergence is that $k_{max} \gamma \gg \omega_{s}$, where $\omega_s$ is the characteristic frequency of the system.\cite{suessHierarchyStochasticPure2014} Because the full auxiliary basis scales as $N_{pig}+k_{max} \choose k_{max}$, it is often impractical to systematically check convergence for sufficiently large values of $k_{max}$. Though our adHOPS method was inspired by localization, we find that it naturally incorporates a dynamic filtering scheme that dramatically improves the scaling of the auxiliary basis with $k_{max}$ even when the exciton is fully delocalized. Fig. \ref{fig:adHOPS_Vscan}a compares the full (black) and adaptive (green) HOPS dynamics with increasing coupling (V). By $V=5\lambda$ the oscillations in the site 3 population report a wave function that is coherently oscillating across 5 sites. Fig. \ref{fig:adHOPS_Vscan}b shows the corresponding size of the auxiliary basis as a function of $k_{max}$. In all cases the adaptive auxiliary basis (green line) increases much more slowly than the full auxiliary basis (black line).

\begin{figure}
  \centering
    \includegraphics{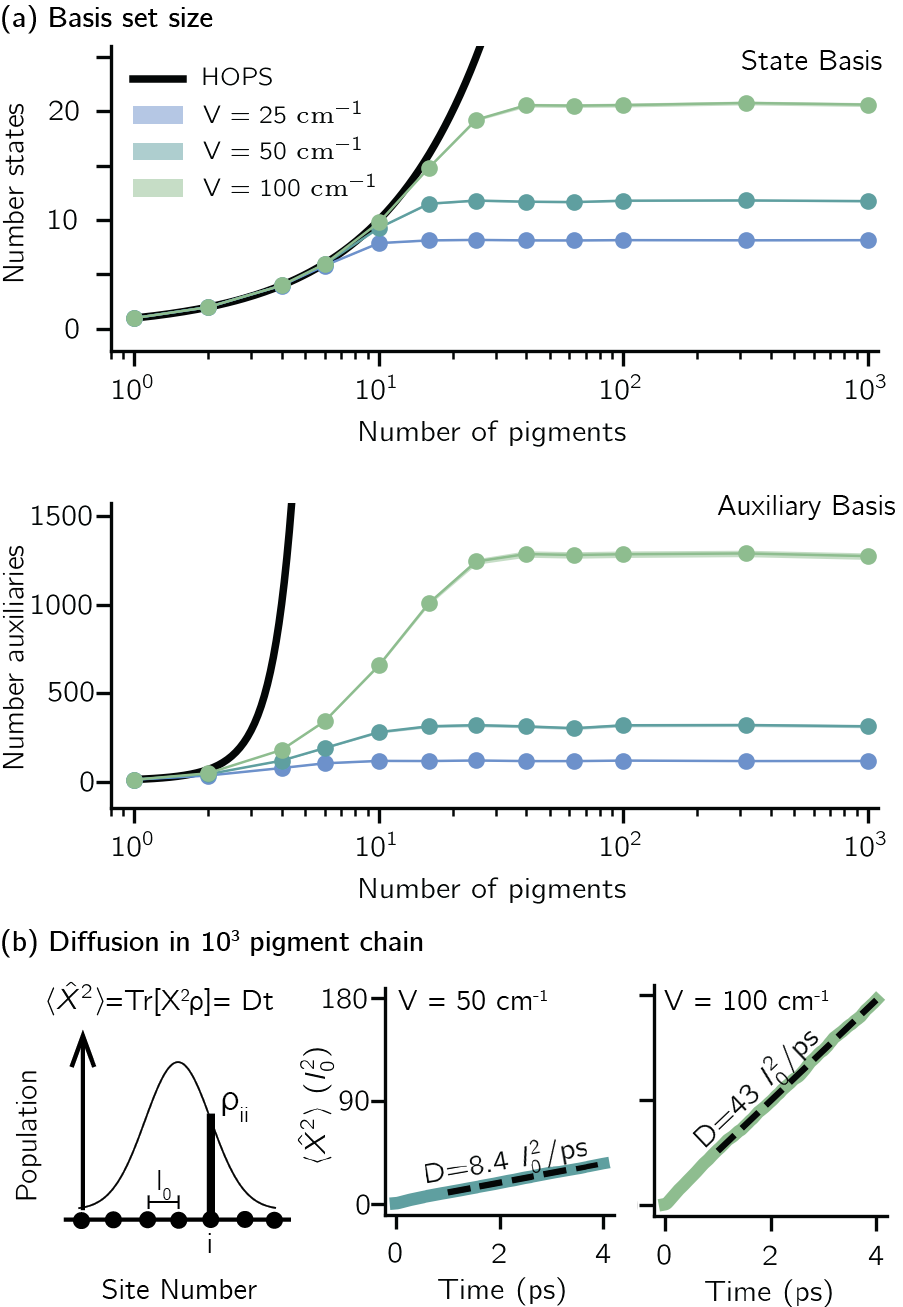}
    \caption{Advantageous scaling of adHOPS simulations for large numbers of pigments. \textbf{(a)} Average number of basis elements for the adaptive system (top) and hierarchy (bottom) for linear chains of different lengths. For V = 25 and 50 cm$^{-1}$ $N_{\textrm{traj}} = 10^3$ and for V = 100 cm$^{-1}$ $N_{\textrm{traj}} = 5 \times 10^3$. \textbf{(b)} Exciton diffusion coefficient (in units of molecular spacing, $l_0$) for a $10^3$ pigment chain from a linear fit to the mean-squared displacement of the excitation ($\textrm{Tr}[\rho \hat{X}^2]$). The excitation starts on the middle pigment (site number is 0). For V= 50 cm$^{-1}$ $N_{\textrm{traj}} = 10^4$ and for V= 100 cm$^{-1}$ $N_{\textrm{traj}} = 5 \times 10^3$. Parameters: $\lambda = \gamma = 50$ cm$^{-1}$, $T$ = 295 K, $k_{max} = 10$, and $\delta = 3\times 10^{-4}$. \label{fig:adHOPS_scaling}}
\end{figure}

Another perpetual challenge for formally exact methods is their intractable computational scaling with the number of molecules. In HOPS calculations this arises from the scaling of the auxiliary basis. Fig. \ref{fig:adHOPS_scaling}a compares the full (black line) size of the state (top) and auxiliary (bottom) basis to the average size of the adaptive basis (colored lines) as a function of the number of molecules in a linear chain. The size of the average auxiliary basis for adHOPS calculations increases much more slowly with the number of molecules than the full auxiliary basis. Moreover, both the auxiliary and state bases in adHOPS calculations show a plateau beyond a threshold size of the linear chain ($N_{pig} > N^*$), indicating the onset of size invariant scaling. In the SI, we compare the CPU time required for full and adaptive HOPS calculations ($V = 50 \textrm{ cm}^{-1}$). We find that adaptive calculations are faster than full calculations starting around $N_{pig} = 10$, and we also demonstrate the onset of size invariance (i.e. $\mathcal{O}(1)$) scaling of CPU time for large aggregates. In other words, increasing the number of pigments beyond a threshold size does not increase the computational expense of an adHOPS calculation. Thus, for localized excitons, the size invariance of adHOPS allows for calculations on scales that were previously unachievable for formally exact methods.

Our adaptive HOPS algorithm offers a computationally tractable approach for formally exact calculations of mesoscale quantum dynamics. As a proof-of-concept, we demonstrate the ability to simulate exciton diffusion on a linear chain of $10^3$ molecules, within the formally exact framework of adHOPS (Fig. \ref{fig:adHOPS_scaling}b). Exciton diffusion is a common experimental observable extracted from non-linear microscopies \cite{ginsbergSpatiallyResolvedExciton2020} but is challenging to simulate on long length scales. \cite{saikinLongRangeExcitonTransport2017,delorCarrierDiffusionLengths2020,wanCooperativeSingletTriplet2015} Using adHOPS, simulating exciton diffusion in a linear chain of $10^3$ pigments is computationally tractable because for V=100 cm$^{-1}$ it requires, on average, less than $2\times 10^3$ auxiliary wave functions and 20 pigment states. The corresponding HOPS simulation would require a auxiliary basis containing $10^{23}$ auxiliary wave functions.

\section{Conclusions}
To summarize, our adaptive HOPS (adHOPS) algorithm: 
\begin{enumerate}
    \item is a formally exact solution to the time evolution of a quantum state coupled to a non-Markovian thermal reservoir,
    \item is embarrassingly (or `perfectly') parallel \cite{herlihy2020art}, and
    \item achieves size-invariant (i.e. $\mathcal{O}(1)$) scaling for large molecular aggregates.
\end{enumerate} 
This combination of properties allows us to perform non-perturbative, non-Markovian simulations involving an arbitrary number of pigments in physically relevant parameter regimes, thus laying the foundation for mesoscale quantum dynamics simulations of excited-state carriers in molecular materials. Future work to extend our adaptive algorithm will allow adHOPS calculations for a broader class of mechanisms involving high-frequency intra-molecular vibrations \cite{beraImpactQuantizedVibrations2015a, blauLocalProteinSolvation2018, bennettMechanisticRegimesVibronic2018} and Peierls-type electron-vibration coupling.\cite{nematiaramModelingChargeTransport2020} Looking forward, we think adHOPS provides a promising new direction for simulations of a broad range of organic semiconductors including photosynthetic membranes,\cite{amarnathMultiscaleModelLight2016, macgregor-chatwinLateralSegregationPhotosystem2017} molecular thin films,\cite{ariasThermallyLimitedExcitonDelocalization2013, sharifzadehLowEnergyChargeTransferExcitons2013} and organic photovoltaic heterojunctions.\cite{hoodEntropyDisorderEnable2016, whaleyCoherentDiffusiveTime2014, jailaubekovHotChargetransferExcitons2013, monahanDirectObservationEntropyDriven2015}

\section*{Supplementary Material}
See supplementary information for the detailed convergence calculations, description of error distribution in the adHOPS trajectory ensemble, and for CPU timing comparisons between HOPS and adHOPS.

\section*{Acknowledgements}
We thank the Robert A. Welch Foundation (Grant N-2026-20200401) and start-up funds from SMU for financial support. DIGB thanks Alex Eisfeld, Richard Hartmann, and Walter Strunz for insightful conversations. 

\section*{Data Availability Statement}
The data that supports the findings of this study, the scripts used to run calculations, and the code required to generate the figures are available at http://doi.org/10.5281/zenodo.4597068. The most recent release of MesoHOPS is available through GitHub at https://github.com/MesoscienceLab/mesohops. The source code used for these calculations is available at http://doi.org/10.5281/zenodo.4592583. 


%
%

%


\bibliography{adHOPS.bib}

\end{document}